\documentclass[12pt]{iopart}
\usepackage{float,graphicx}
\usepackage{psfig,verbatim}

\begin{document}

\title{Barkhausen noise from {\it zigzag} domain walls}


\author{B. Cerruti$^1$ and S. Zapperi$^1$}

\address{$^1$CNR-INFM, Dipartimento di Fisica, Universit\`a ``La Sapienza'', P.le A. Moro 2, 00185 Roma, Italy}

\ead{\mailto{benedetta.cerruti@gmail.com}, \mailto{stefano.zapperi@roma1.infn.it}}

\begin{abstract}
We investigate the Barkhausen noise in ferromagnetic thin films with {\it zigzag} domain walls. We use a cellular automaton model that describes the motion of a {\it zigzag} domain wall in an impure ferromagnetic quasi-two dimensional sample with in-plane uniaxial magnetization at zero temperature, driven by an external magnetic field. The main ingredients of this model are the dipolar spin-spin interactions and the anisotropy energy. A power law behavior with a cutoff is found for the probability distributions of size, duration and correlation length of the Barkhausen avalanches, and the critical exponents are in agreement with the available experiments. The link between the size and the duration of the avalanches is analyzed too, and a power law behavior is found for the average size of an avalanche as a function of its duration.
\end{abstract}

\maketitle

\section{Introduction}
Understanding the properties of ferromagnetic thin films 
is still one of the open questions in the physics of magnetic systems. An 
important issue is linking the domains and domain walls structure 
of a material with its hysteretic
properties, like the Barkhausen noise (BN).
The Barkhausen effect \cite{barkhausen} 
consists in the irregularity of the magnetization variation while 
magnetizing a sample with a slowly varying 
external magnetic field, and is due to the 
jerky motion of the domain walls in a system with structural disorder 
and impurities. 
Once the origin of BN was understood \cite{wisho}, it was soon 
realized that it could be used as an effective probe to investigate 
the magnetization dynamics in magnetic materials.
Furthermore, from a purely theoretical point of view, it is a 
good example of dynamical critical behavior, as evidenced by experimental 
observation of power law distributions for the statistics 
of the avalanche size and duration \cite{sethna}. 
Moreover, there is a growing evidence that soft magnetic bulk materials can
be grouped in different classes according to the scaling 
exponents values \cite{dz},
so that BN could be seen as a non destructive experimental 
tool for the analysis of the properties of a material, and a similar feature 
could be expected to hold for two-dimensional materials.  

Until now, most of the models and experiments on BN have focused on three 
dimensional systems \cite{BK}. 
The difficulties for the study of two dimensional 
samples, in fact, concern both the theoretical and the experimental 
aspects of the problem.
On the theoretical side, the topology of domain and domain walls is 
much more rich and complicated in two dimension (parallel or head-on domains, 
charged and 
uncharged walls, magnetization in or out the film plane, parallel or 
{\it zigzag} walls, labyrinthic domains, etc.) than in the bulk case
\cite{hubert} (mainly parallel domains with uncharged walls), so it is not 
obvious how to generalize the well known three dimensional models.
The models currently used for three dimensional materials 
could be classified in two main groups, namely spin models of the Ising
type \cite{RFIM,vives,vives2,vives3}, and single domain wall models
\cite{zapperi,zapperi2,urbach,narayan,bahiana,queiroz}. 
On the experimental side, the
inductive experimental setup commonly used for bulk samples \cite{BK} is 
usually not suitable for thin films, due to the low 
intensity of the magnetic flux variation signal, 
which is roughly proportional to the 
sample thickness, and thus tends to vanish for very thin films. 
Conversely, the magneto-optical \cite{walsh,pupprl,schwarz} and 
magneto-optical microscope magnetometry
\cite{kim} experiments, though able to supply
the domain structure, that is not accessible by the inductive measurements, 
provide only partial informations about the probed zone, 
and not about the whole sample. 
So, despite the increasing interest concerning ferromagnetic thin films
applications in magnetic recording technology and spintronic devices, 
a complete understanding of two dimensional 
system behavior is still lacking.

In this article we focus on two dimensional systems with 
{\it zigzag} domain walls, which arise 
in thin films with head-on magnetization between nearest-neighbor domains,
mainly due to the balance between the magnetostatic 
and the anisotropy contributions to the system total energy \cite{freiser}. 
Those kind of walls have been observed for the first time 
in thin film magnetic recording media,
where head-on domains are induced by means of the application of a recording
head field, and have been then observed in films of several magnetic media
such as iron \cite{iron}, Co \cite{co}, Gd-Co \cite{freiser}, 
epitaxial Fe films grown on GaAs(001) \cite{bland1},
ferrite-garnet films with strong cubic anisotropy 
\cite{vlasko} and many others. 
This kind of walls have been observed too in ferroelectric materials,
such as Gd$_2$(MoO$_4$)$_3$ crystals \cite{alexeyev}.

We present a study of the Barkhausen noise at $T=0$.
We use a slightly modified version of a simple single-wall 
discrete model for the 
motion of the {\it zigzag} wall that we recently proposed 
for the study of the dynamic hysteresis in ferromagnetic thin 
films \cite{benedetta}. 
Our model is based on the interplay between dipolar and anisotropy 
energy contributions, in the presence of structural disorder and external 
magnetic field. Via cellular automaton simulations 
the model describes the motion, in a disordered 
ferromagnetic thin film, of a {\it zigzag} wall between two domains of 
opposite magnetizations meeting head-on at the wall,
 up to the saturation of the magnetization driven by the
external magnetic field.
We find that the probability distributions of  
the size, the duration and the correlation length of the Barkhausen avalanches
show a power law behavior with a cutoff, 
and the critical exponents are in 
quantitative agreement with experimental data two dimensional sample, 
as Co polycristalline thin films \cite{kim,kim2}.
Finally, in order to have a deeper insight on the link between 
the size and the 
duration of the avalanches, we study the behavior of the average size of an
avalanches as a function of its time duration, and find that it could 
be described too by a power law.

\section{The model}

\begin{figure}[h]
\centerline{\psfig{figure=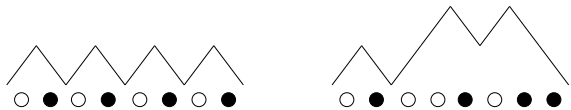,width=7.5cm,clip=}}
\caption{The mapping. Two examples of {\it zigzag} configurations. A segment with a positive (negative) slope corresponds to a void (particle).}
\label{zigzag}
\end{figure}

Our purpose is to study the motion of a single domain wall 
driven by an external magnetic field,
by discrete model simulations. To this end, 
we start from a model that we have recently introduced \cite{benedetta}:
as we are interested in the macroscopic response,
the aim of our model is to discretize the {\it zigzag} wall in minimum
segments in order to map the quasi-two dimensional problem of the wall
motion in a one dimensional model (Fig. \ref{zigzag}), regardless
the details 
of the wall internal structure that are not expected to influence
the macroscopic length scale.

We calculate the total energy of an arbitrary 
{\it zigzag} wall configuration, considering only
the magnetostatic, the anisotropy and the disorder contributions, 
and the interaction with the external magnetic field:
\begin{equation}
E=E_m+E_{an}+E_{dis}+E_{ext}\mbox{.}
\label{E}
\end{equation}
In the equation (\ref{E}), 
the magnetostatic term $E_m$ takes into account the interaction between 
the magnetization and the stray field, the anisotropy $E_{an}$ is the
energy cost of the deviations from the easy axis of the material, that
are associated with the N\'eel tail surrounding the wall \cite{benedetta}, 
and $E_{dis}$ models
structural disorder, impurities, defects etc.. 

\begin{figure}[h]
\centerline{\psfig{figure=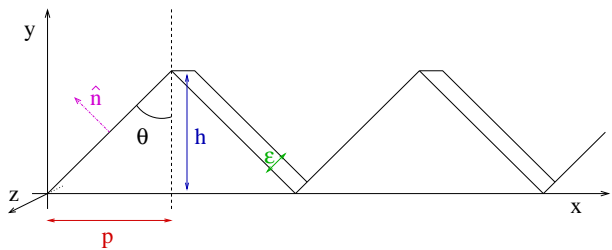,width=8.cm,clip=}}
\caption{Sketch of the parameters of the {\it zigzag} wall. The easy axis is
along the $y$ direction.}
\label{sample}
\end{figure}

As we have discussed in our previous paper \cite{benedetta}, the magnetostatic
interaction energy between two segments $i$ and $j$ could be 
approximated as
$$E_{m,ij}=8M_s^2\epsilon^2p^2\mu_0\frac{1}{r_{ij}}\mbox{,}$$
where $M_s$ is the saturation magnetization, $\epsilon$ is the sample 
thickness, $p$ is the minimal half-period of a {\it zigzag} configuration,
$\mu_0$ is the vacuum permeability and $r_{ij}$ is the distance between the 
centers of mass of the two segments $i$ and $j$ (see Fig. \ref{sample}).

\vspace{1cm}

The anisotropy energy term 
in the simple case of an uniaxial crystal can be written as
\begin{equation}
E_{an}=\int\!\!\! d^3r K_u \sin^2\phi  ,
\label{anisotropy}
\end{equation}
where $K_u$ is the in-plane 
uniaxial anisotropy constant and $\phi$ is the angle 
between the easy axis and the magnetization vector.
Assuming \cite{freiser} that the magnetic charge 
associated with the magnetization rotation is 
uniformly distributed over the entire band containing the wall and a
linear in-plane rotation of the 
magnetization vector,  
we obtain the anisotropy energy for unit 
length 
\begin{equation}
E_{an}=\epsilon K_u h c(\theta) ,
\label{E_an}
\end{equation}
where $K_u$ is the anisotropy constant of the material, $h$ is the 
{\it zigzag} amplitude and 
$c(\theta)$ is a constant function of the {\it zigzag} angle $\theta$
which could be evaluated numerically (see \cite{benedetta} for more details).

Concerning the $E_{dis}$ term, we consider only quenched (frozen)
disorder, that does not evolve on the timescale of the magnetization
reversal.  We model the disorder by an energy contribution 
associated to each site of our discretized sample which may be occupied by a segment
(our discrete unit length) of the {\it zigzag} wall. This
term is extracted from an uncorrelated random Gaussian distribution with zero
mean.

The interaction energy with the external magnetic field $H_{ext}$, that is set on
the easy axis direction, is given by
\begin{equation}
E_{ext}=-\mu_0 H_{ext}M ,
\label{Eext}
\end{equation}
where $M$ is the magnetization of the system.

The dynamics of the model is implemented by 
switching on the external magnetic field $H_{ext}$,
looking for all the pairs of segments with slope
up-down in the wall, i.e. the void-particle pairs,
and trying to exchange the positions in the pair.
This rule allows only the forward motion 
of pairs of segments with up-down slope, and preserves 
the {\it zigzag} (solid-on-solid) structure of the wall.
Once a possible displacement has been attempted, we calculate the total energy 
difference $\Delta E$ between
the starting configuration and the new one by using Eq. (\ref{E}).
The move is accepted if $\Delta E\le 0$, 
and in that case we update the configuration and continue the process,
otherwise we reject it.
The acceptance of the move corresponds to the starting of an avalanche, 
that goes on until the wall comes to rest,
i.e. when the minimum difference of energy $\Delta E_{min}$ 
over all the void-particle pairs is
bigger than zero. 
To restart the process and eventually trigger another avalanche, 
we increase the external field by an amount 
$\Delta H_{ext}=\Delta E_{min}/\mu_0 \Delta M$, where $\Delta M$ 
is the variation of the magnetization due to the flip of the void-particle 
pair, and continue the updating. For this parallel dynamics, we can 
identify the number of the iterations of the updating with 
the physical time. We begin the simulation from the 
$M=0$ at $H_{ext}=0$ state and drive the sample to the
positive saturation (our sample has a finite size).  

\section{Results}

\begin{figure}[h]
\centerline{\psfig{figure=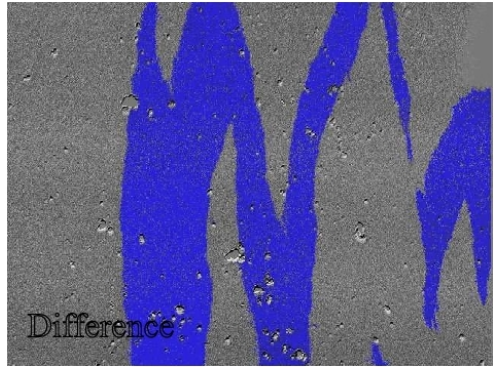,width=6.5cm,clip=}}
\caption{Magneto-optical experimental data: the colored area represents the spatial difference between the two configurations before and after an avalanche (i.e. the size of the avalanche). Courtesy of G. Durin.}
\label{diff_sper}
\end{figure}

\begin{figure}[h]
\centerline{\psfig{figure=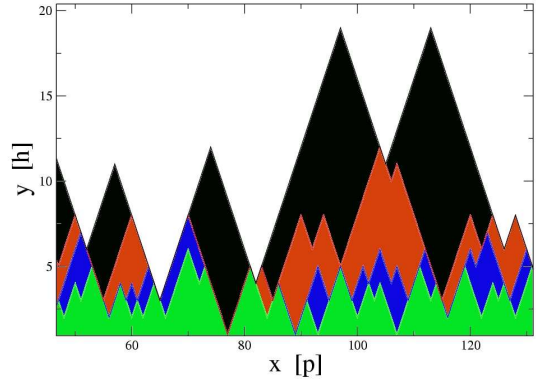,width=8.2cm,clip=}}
\caption{A zoom of four avalanches in a simulation. The colored areas represents the difference of magnetization between two successive configurations before and after an avalanche (in units of half-period and amplitude of the {\it zigzag}).}
\label{diff_model}
\end{figure}

From experimental magneto-optical observations (Fig. \ref{diff_sper}) 
we can infer that 
the dynamics of the wall is jerky and proceeds by
avalanches, preserving the value of the {\it zigzag} angle and with a 
gradual increase of the {\it zigzag} period and amplitude. These features are
recovered in our simulations (see Fig. \ref{diff_model}): the wall may be
pinned by the disorder and the depinning of a void-particle pair driven
by the external magnetic field could
eventually trigger an avalanche. Moreover, when the field increases, 
the anisotropy and disorder energy terms become less important compared
with $E_{ext}$. So the relevance of the magnetostatic
interactions increases, and since this term tends to move the 
magnetostatic charges (all of the same sign at the wall) away from each other, 
it drives a coarsening of the {\it zigzag} segments, 
whose period thus increases.

An avalanche is defined as the event between two pinned configuration
of the domain wall. Since in our model we only allow the forward motion of the 
wall, we can thus identify univocally the time duration and 
spatial size of an avalanche:
the time duration is the number of updates from the 
depinning of the wall to the new pinned configuration, and 
the size is defined as the area interested by the 
magnetization reversal during an avalanche. 
It is thus possible to analyze the statistics of the 
avalanches, by constructing the probability distributions of the size $S$,
the duration $T$ and the correlation length $\xi$.

\begin{figure}[h]
\begin{center}
\begin{tabular}{c c}
\includegraphics[height=8.5cm,angle=-90,clip=]{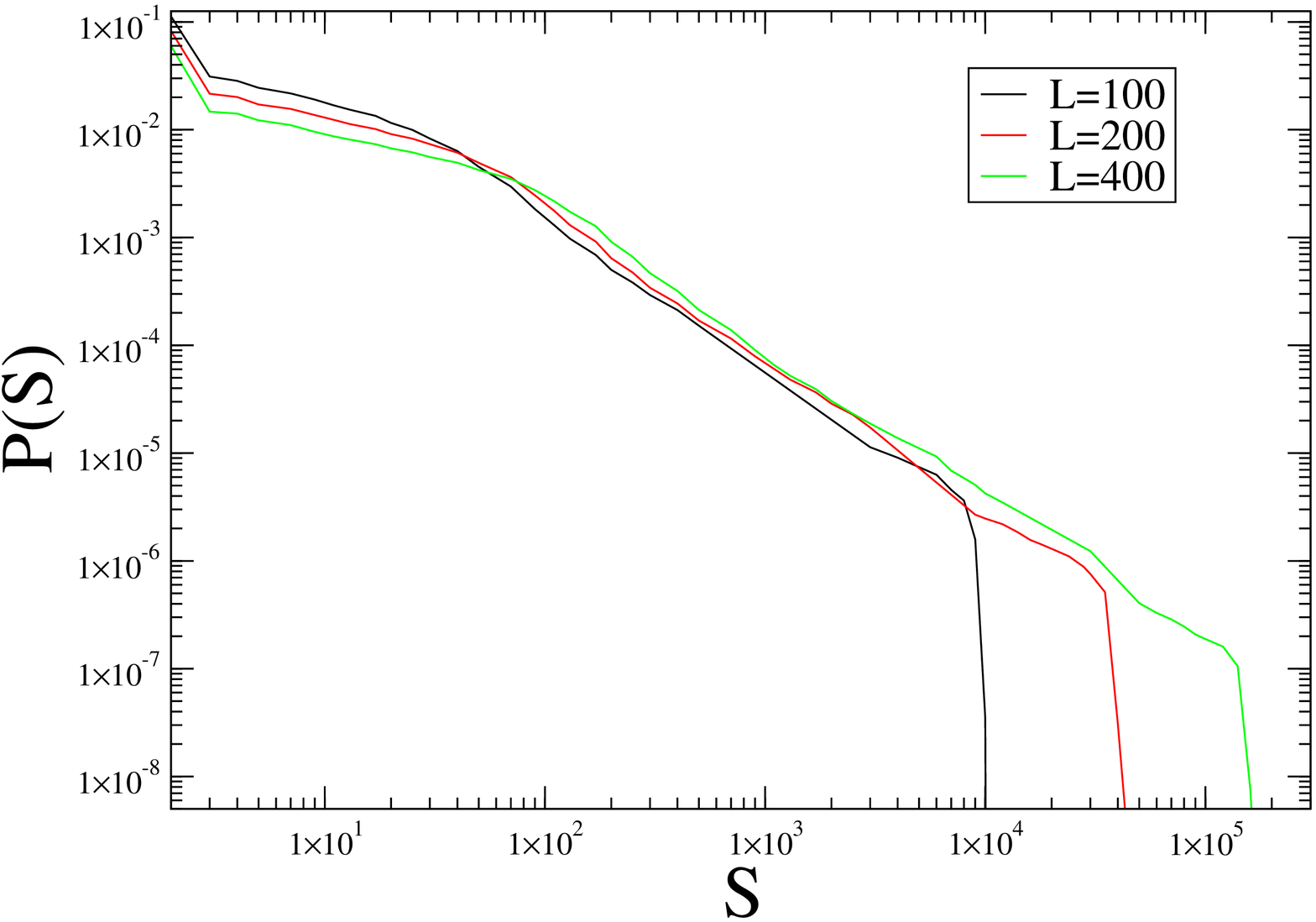} &
\includegraphics[height=8.5cm,angle=-90,clip=]{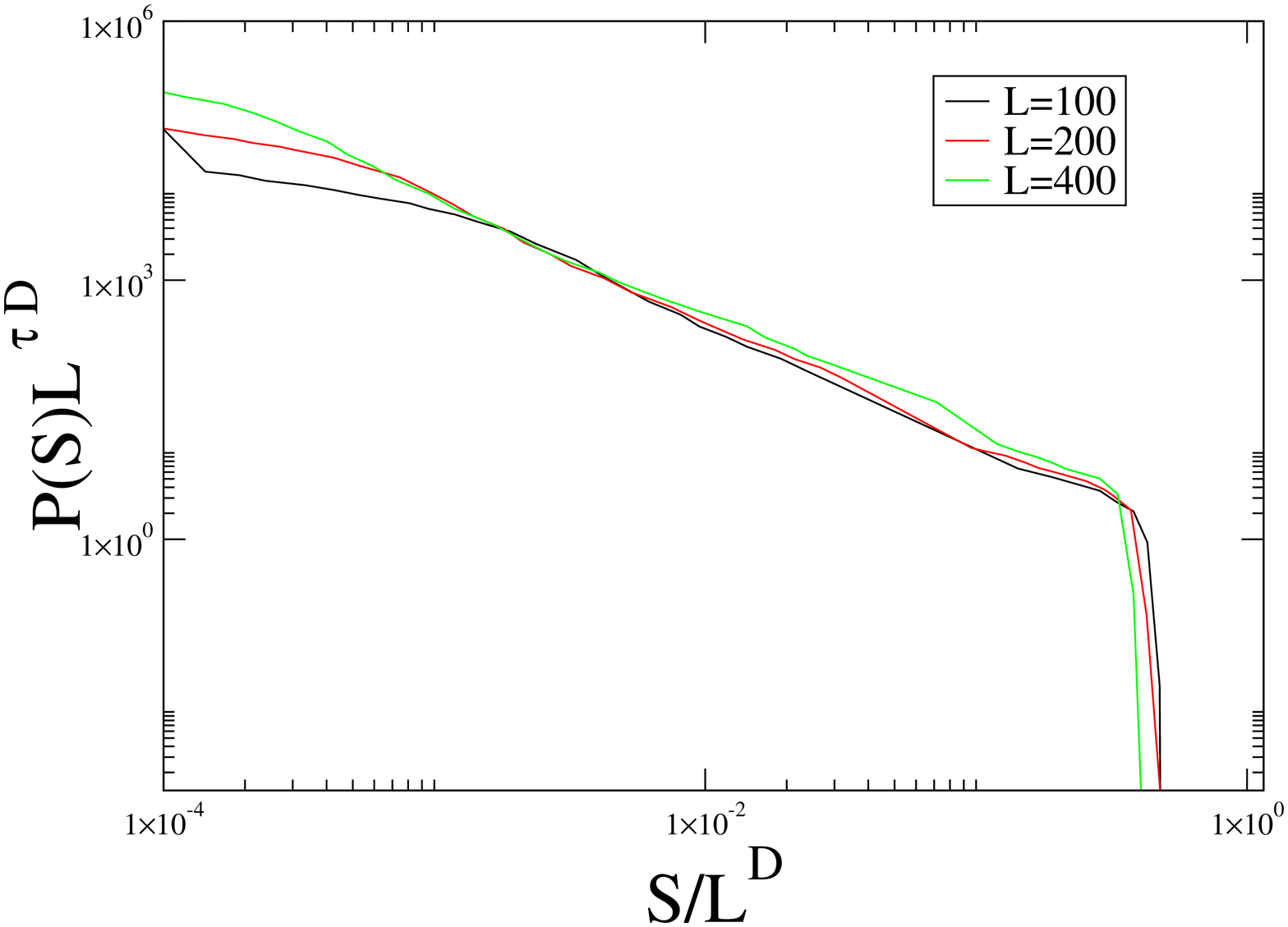} \\
(a)&(b)\\
\end{tabular}
\end{center}
\caption{(a) Probability distribution for the avalanche size $S$, for three different values of the total sample length $L$. (b) The probability distribution for the rescaled variable $S/L^D$, where $D$ is a fitted scaling exponent which value is $2.1611$. In the power law region of the plots, they collapse on the same curve.}
\label{PdiS}
\end{figure}

\begin{figure}[h]
\begin{center}
\begin{tabular}{c c}
\includegraphics[height=8.5cm,angle=-90,clip=]{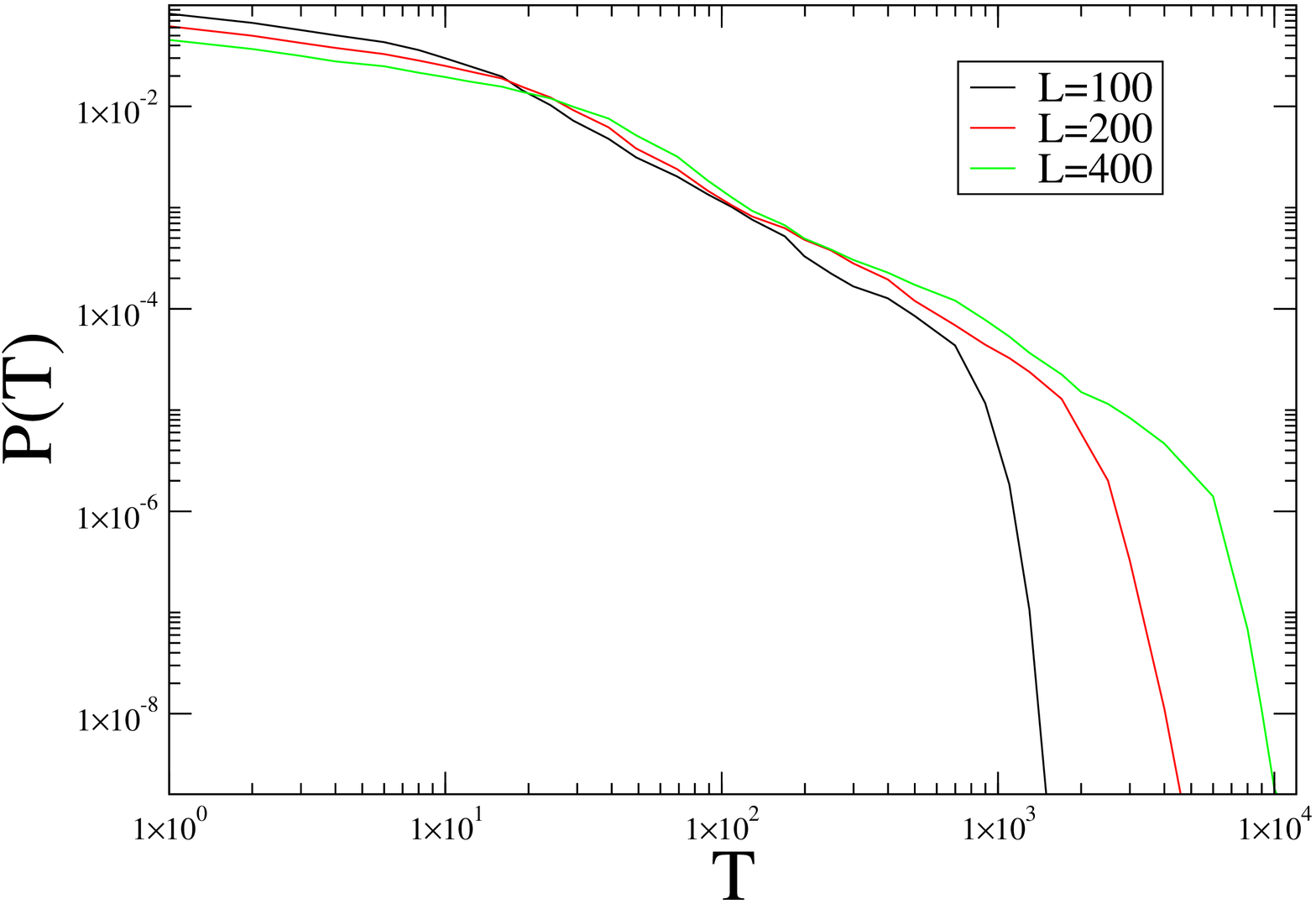} &
\includegraphics[height=8.5cm,angle=-90,clip=]{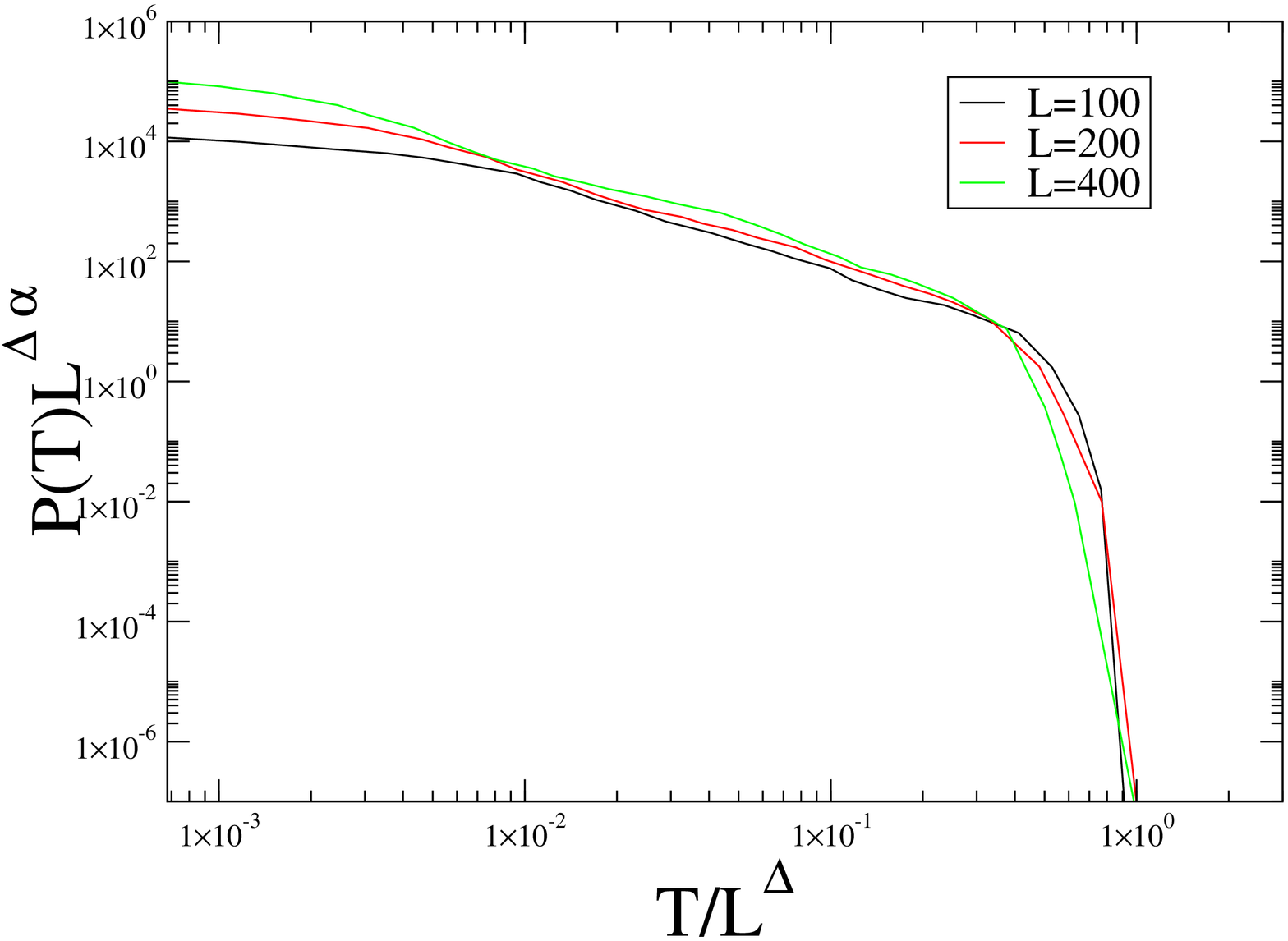} \\
(a)&(b)\\
\end{tabular}
\end{center}
\caption{(a) Probability distribution for the avalanche duration $T$, for three different values of the total sample length $L$. (b) The probability distribution for the rescaled variable $T/L^\Delta$, where $\Delta$ is a fitted scaling exponent which value is $1.6149$. In the power law region of the plots, they collapse on the same curve.}
\label{PdiT}
\end{figure}

As it can be seen from the figures \ref{PdiS} and \ref{PdiT}, 
both the distributions
show a power law behavior for almost two decades,
respectively $P(S)\sim S^{-\tau}f(S/S_0)$ for the size and
$P(T)\sim T^{-\alpha}g(T/T_0)$ for the time duration  probability function, 
and a cutoff, which values are $S_0$ and $T_0$. The nature of the 
cutoff, which is an experimentally well known feature, is in our model due to
finite size effects, as it could be seen comparing the distributions for 
different number of {\it zigzag} minimum segments $n$. 
We find that the cutoff distributions scale respectively as
$S_0\sim L^D$ and $T_0\sim L^\Delta$, where $L$ is the total 
length of the sample, $L=np$, and with $D\sim 2.16$ and $\Delta \sim 1.61$.
In Fig. \ref{PdiS}$(b)$ and \ref{PdiT}$(b)$ we show that rescaling 
the variables as $S/L^D$ and $T/L^{\Delta}$, the rescaled 
probability distributions $P(S)L^{\tau D}$ and $P(T)L^{\alpha \Delta}$
obtained for the various system sizes, collapse on the same curve.
The power law behavior is the 
fingerprint of a dynamical critical behavior, and could be characterized by 
the associated critical exponents. For our model, the critical exponents 
for the avalanches size and the duration distributions are
respectively $\tau\sim 1.34$ and $\alpha\sim 1.55$, and do not essentially 
depend on the system size, neither on the intensity of the 
anisotropy and the disorder, 
that may beside influence the very low $S$ or $T$ zones of the probability 
distributions. The value of $\tau$ is in good
agreement with recent 
experimental results on Co polycristalline thin films 
\cite{kim,kim2}, that give $\tau\sim 4/3$, 
although this reference does not provide informations on the duration
statistics. In these references \cite{kim,kim2}, 
the authors report direct, time-resolved
domain observations of a Barkhausen avalanche obtained by a magneto-optical
microscope magnetometer, which directly visualizes the microscopic behavior 
of the avalanches in thin films. The observations 
are restricted to a limited zone of the sample for experimental reasons, 
and domain topology shows a system of tips, that could be interpreted as a 
a part of a {\it zigzag} domain wall.
Anyway, more experimental confirmations would be necessary.

\begin{figure}[h]
\centerline{\psfig{figure=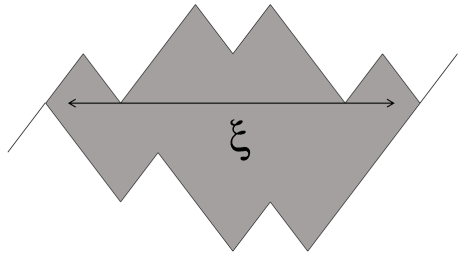,width=6.cm,clip=}}
\caption{Definition of the correlation length in an avalanche. The gray zone represents the area interested by the avalanche.}
\label{valanga_zigzag}
\end{figure}

From our simulations we can even study the statistics of the correlation length
$\xi$ of the avalanches. $\xi$ is defined as the longitudinal size of the 
avalanche (see Fig. \ref{valanga_zigzag}), and gives a measure of the 
portion of the wall interested by the avalanche itself.

\begin{figure}[h]
\begin{center}
\begin{tabular}{c c}
\includegraphics[height=8.5cm,angle=-90,clip=]{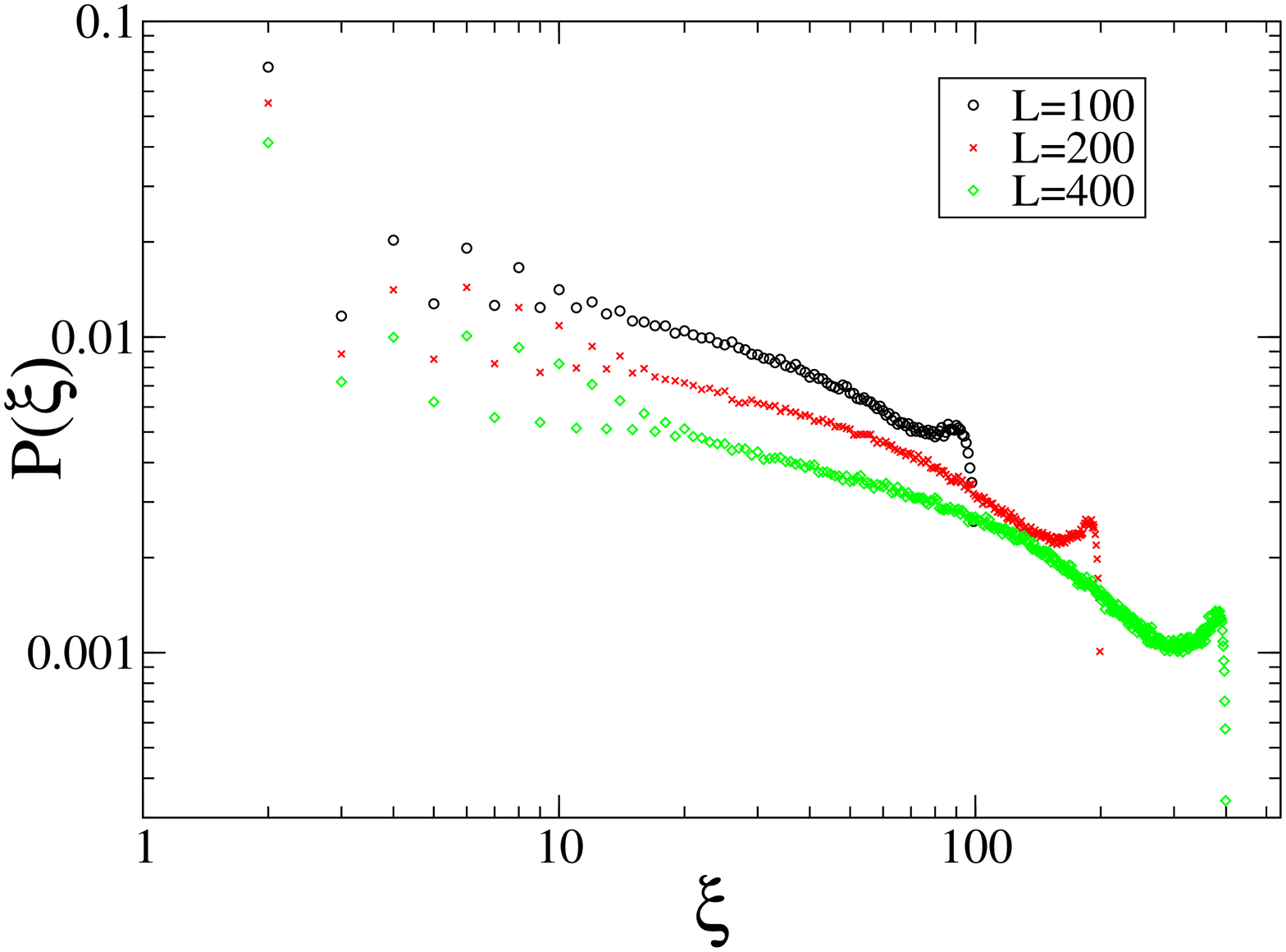} &
\includegraphics[height=8.5cm,angle=-90,clip=]{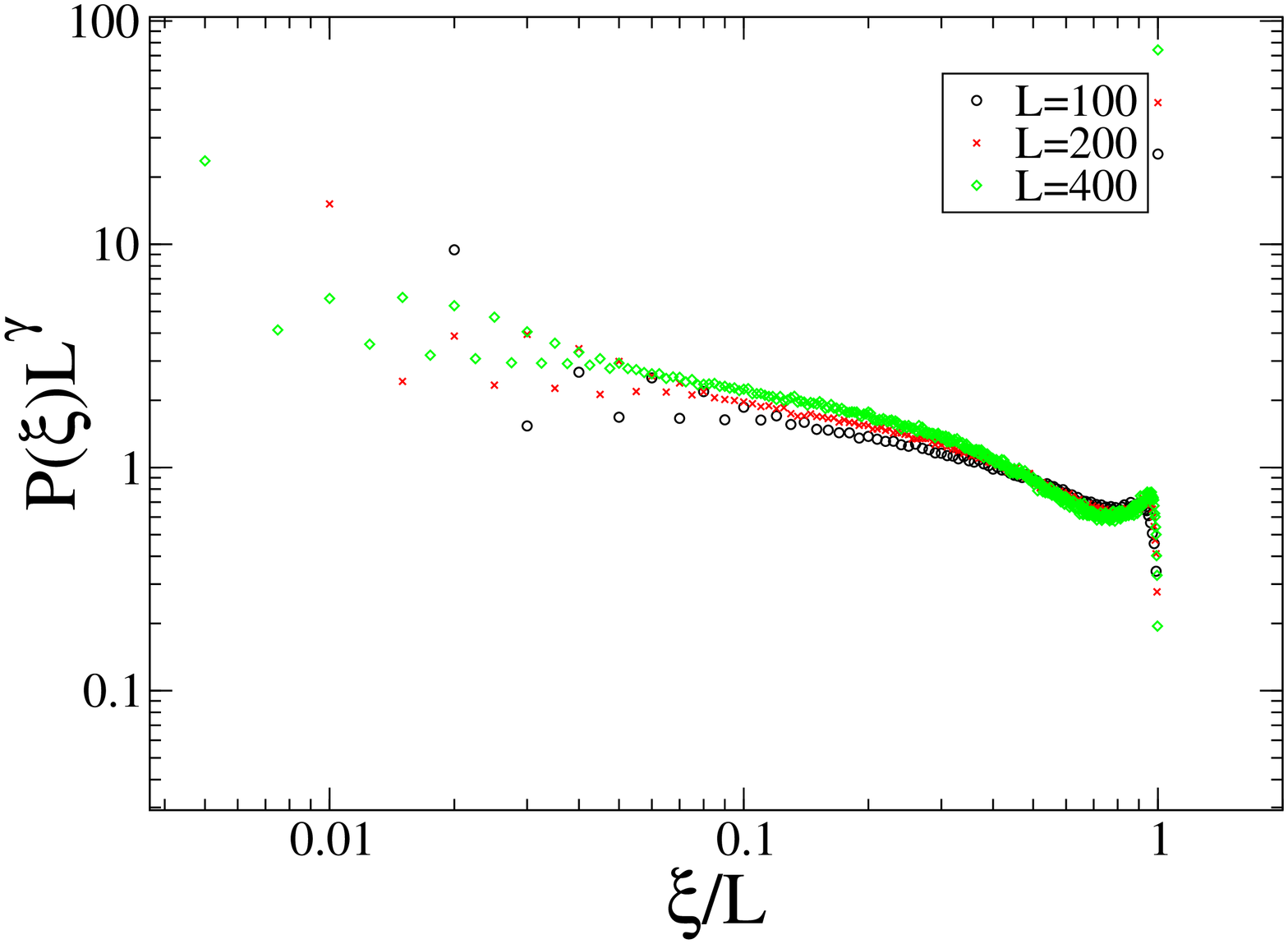} \\
(a)&(b)\\
\end{tabular}
\end{center}
\caption{(a) Probability distribution for the avalanche correlation length (in unit of the minimum {\it zigzag} half period $p$, for three different values of the total sample length $L$. (b) The probability distribution for the rescaled correlation length $\xi/L$. In the power law region of the plots, they collapse on the same curve.}
\label{lunghezza_correlazione}
\end{figure}

The probability distribution of the correlation lengths follows 
too a power law 
behavior, $P(\xi)\sim \xi^{\beta} h(\xi/\xi_0)$ where $\xi_0$ is the cutoff
 (Fig. \ref{lunghezza_correlazione}), with an exponent close to $-0.4$.
The low $\xi$ zone of Fig. \ref{lunghezza_correlazione} is very noisy because 
of the topology of the model: small avalanches with a correlation length 
of an even number of minimal segment are more probable than
avalanches with odd $\xi$, 
because for the correlation length study we consider the sum of the
events in one uploading step, even if they are not geometrically connected. 
So is more likely to have, for instance, two elementary events in two separate
parts of the sample than two connected events, for small avalanches size.
Moreover, for very high values of $\xi$, we notice another deviation from
the power law behavior, this time due to the finite size of the sample. In fact
the avalanches that span the whole sample could not have a correlation length 
bigger than the sample size, so the probability distributions show peaks 
for values of $\xi$ around 
the system size, which are just an artifact of finite simulations, 
as of course 
the cutoff at the system length value.
The cutoff distribution scales as $\xi_0\sim L$. 
In Fig. \ref{lunghezza_correlazione}$(b)$ we show that rescaling 
the correlation length as $\xi/L$, the rescaled 
probability distributions $P(\xi)L^\beta$ obtained 
for the various system sizes collapse on the same curve.

\begin{figure}[h]
\centerline{\psfig{figure=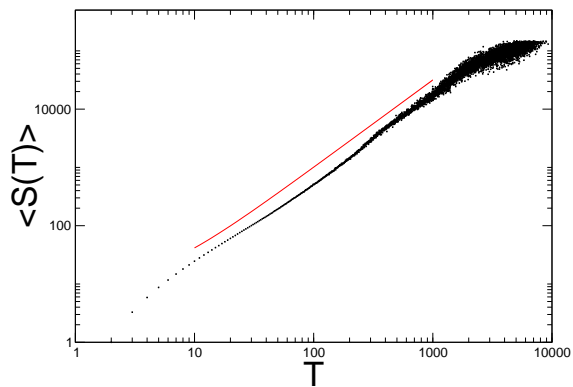,width=8.2cm,angle=-90,clip=}}
\caption{Average size of an avalanche vs its duration $T$ for $L=400$. A part form the cutoff at large values $T$ and effects of the discretization of the model for small $T$, the function shows a power law behavior for almost two decades with an exponent close to $\gamma\sim 1.5$ (solid line is a power law with exponent $1.45$ and is a guideline to the eye).}
\label{smedio_vs_t}
\end{figure}

Another interesting issue that could be studied is the correlation between the size
and the duration of the avalanches. Since avalanches with the same duration could
show quite different sizes, this feature could be quantified by 
addressing to the link
between the mean size $<\!\!\!S(T)\!\!\!>$ of an avalanche and its
duration $T$ (Fig. \ref{smedio_vs_t}). 
This function follows a power law behavior 
$<\!\!\!S(T)\!\!\!>\sim T^{\gamma}$ with an 
exponent close to $\gamma\sim 1.5$ 
for all the three sample sizes that we have investigated 
($L=100$, $L=200$ and $L=400$). 

We can check the consistency of the 
exponent $\gamma$ considering that it 
must be $P(S)dS\sim P(T)dT$. Using the power law $P(S)\sim S^{-\tau}$,
$P(T)\sim T^{-\alpha}$ and $<\!\!\!S(T)\!\!\!>\sim T^{\gamma}$,
it can be easily derived that it must be 

$$\alpha=\gamma(\tau-1)+1 .$$

Using $\gamma=1.45$ and $\tau=1.34$ we would
obtain $\alpha=1.49$ that is in reasonable agreement with the measured
value $\alpha=1.55$.

\section{Conclusions}
The Barkhausen noise is known to be due to the jerky motion of the domain walls
in a disordered material during the magnetization process. Even 
if the essential physics of the problem is well understood, many questions are 
still open. One of the most challenging related 
topics is the physics of ferromagnetic thin films, 
which displays new features with respect to bulk materials, 
and still remains to be
fully explored both experimentally and theoretically.
In this work we have applied a slightly modified version of a model that we 
have recently proposed \cite{benedetta} 
for the study of dynamic hysteresis for systems with 
{\it zigzag} domain walls. This model takes into account the contribution to the
total energy of the dipolar long range interactions, the anisotropy and the
disorder. The dynamics of the model describes qualitatively the experimentally 
observed features of the domain wall motion, like the jerky nature of the 
motion and the coarsening of the {\it zigzag} amplitude. We have studied 
the size, the duration and
the correlation length of the avalanches by means of their 
probability distributions, via cellular automaton simulations. 
All these distributions show a power law behavior 
and a cutoff due to finite size effects. 
The critical exponents of the three distributions are derived and we find a good 
agreement for the value of $\tau\sim 1.34$, associated with the size statistics, 
with recent magneto-optical 
measurements on Co polycristalline thin films \cite{kim}. Anyway, more
experimental confirmations are needed, especially for the duration and 
the correlation length distributions, for which no experimental data are available
up to now. Finally, we have investigated the link between the size and the 
duration of the Barkhausen avalanches, by studying the average size of 
an avalanche as a function of its duration $T$. 
Even this function follows a power law,
with a critical exponent close to $1.5$.

\end{document}